\documentclass[dvips]{acta}
\usepackage{supertabular,lscape,epsfig}
\usepackage{amssymb}
\usepackage{amsmath}
\DeclareSymbolFont{ppa}{OT1}{ppl}{m}{it}
\DeclareMathSymbol{\vv}{\mathalpha}{ppa}{'166}

\newfont{\hb}{rphvb at 10pt}
\newfont{\hbo}{rphvbo at 10pt}
\newfont{\bitt}{rptmbi at 12pt}
\newfont{\bits}{rptmbi at 11pt}

\SetPages{197}{204}

\SetVol{65}{2015}

\begin{document}

\newcommand{\TabApp}[2]{\begin{center}\parbox[t]{#1}{\centerline{
  {\bf Appendix}}
  \vskip2mm
  \centerline{\small {\spaceskip 2pt plus 1pt minus 1pt T a b l e}
  \refstepcounter{table}\thetable}
  \vskip2mm
  \centerline{\footnotesize #2}}
  \vskip3mm
\end{center}}

\newcommand{\TabCapp}[2]{\begin{center}\parbox[t]{#1}{\centerline{
  \small {\spaceskip 2pt plus 1pt minus 1pt T a b l e}
  \refstepcounter{table}\thetable}
  \vskip2mm
  \centerline{\footnotesize #2}}
  \vskip3mm
\end{center}}

\newcommand{\TTabCap}[3]{\begin{center}\parbox[t]{#1}{\centerline{
  \small {\spaceskip 2pt plus 1pt minus 1pt T a b l e}
  \refstepcounter{table}\thetable}
  \vskip2mm
  \centerline{\footnotesize #2}
  \centerline{\footnotesize #3}}
  \vskip1mm
\end{center}}

\newcommand{\MakeTableApp}[4]{\begin{table}[p]\TabApp{#2}{#3}
  \begin{center} \TableFont \begin{tabular}{#1} #4 
  \end{tabular}\end{center}\end{table}}

\newcommand{\MakeTableSepp}[4]{\begin{table}[p]\TabCapp{#2}{#3}
  \begin{center} \TableFont \begin{tabular}{#1} #4 
  \end{tabular}\end{center}\end{table}}

\newcommand{\MakeTableee}[4]{\begin{table}[htb]\TabCapp{#2}{#3}
  \begin{center} \TableFont \begin{tabular}{#1} #4
  \end{tabular}\end{center}\end{table}}

\newcommand{\MakeTablee}[5]{\begin{table}[htb]\TTabCap{#2}{#3}{#4}
  \begin{center} \TableFont \begin{tabular}{#1} #5 
  \end{tabular}\end{center}\end{table}}

\newfont{\bb}{ptmbi8t at 12pt}
\newfont{\bbb}{cmbxti10}
\newfont{\bbbb}{cmbxti10 at 9pt}
\newcommand{\uprule}{\rule{0pt}{2.5ex}}
\newcommand{\douprule}{\rule[-2ex]{0pt}{4.5ex}}
\newcommand{\dorule}{\rule[-2ex]{0pt}{2ex}}
\begin{Titlepage}
\Title{On the Atmospheric Extinction Reduction Procedure in Multiband Wide-Field Photometric Surveys}

\Author{A.~~Z~a~k~h~a~r~o~v$^1$,~~ A.~~M~i~r~o~n~o~v$^1$,~~ 
A.~~B~i~r~y~u~k~o~v$^{1,2}$,~~ N.~~ K~r~o~u~s~s~a~n~o~v~a$^1$,\\
M.~~P~r~o~k~h~o~r~o~v$^1$,~~ G.~~B~e~s~k~i~n$^{2,3}$,~~ 
S.~~K~a~r~p~o~v$^{2,3}$,~~ S.~~B~o~n~d~a~r$^4$,~~ E.~~I~v~a~n~o~v$^4$,\\
A.~~P~e~r~k~o~v$^{2,4}$~~ and~~ V.~~S~a~s~y~u~k$^2$}
{$^1$Sternberg Astronomical Institute of M.V. Lomonosov Moscow State University, 13~Universitetsky pr., Moscow, 119991 Russia\\
$^2$Kazan Federal University, 18 Kremlevskaya st., Kazan, 420008 Russia\\
$^3$Special Astrophysical Observatory, Karachai-Cherkessia, Nizhnij Arkhyz, 369167 Russia\\
$^4$Precision Systems \& Instruments Corp., 53 Aviamotornaya st., Moscow, 111024 Russia}

\Received{December 11, 2014}
\end{Titlepage}

\Abstract{We propose an improved method for the atmospheric extinction 
reduction within optical photometry. Our method is based on the
simultaneous multicolor observations of photometric standards. Such
data are now available within the modern wide-field sky surveys and
contain a large amount of information about instant atmospheric
conditions. So, it became possible to estimate the extinction
parameters on the basis of a quite short observational dataset and,
hence, to trace the rapid stars twinkling accurately. Having been
developed for a new MiniMegaTORTORa observational system, the
proposed method can be adopted for a wide range of modern
observational programs.}{Techniques: photometric -- Surveys}

\Section{Introduction}
Large-scale photometric surveys (both ground- and space-based) have
recently become a common method for the investigation of astronomical
sources variability and for standardized multiband photometry. Many
of them are inexpensive and easy to be implemented since their design
is based on the usage of small telescopes. At the same time these
surveys deliver a large amount of homogeneous observational data that
can be processed and analyzed automatically. Typical exposure times
are becoming shorter in the course of the years making it possible to
investigate astrophysical phenomena with high temporal resolution.

The examples of such modern multiband optical sky surveys are
large-scale LSST (Ivezic \etal 2008), PanSTARRS (Kaiser \etal 2002),
Gaia (Lindegren and Perryman 1996) and more specialized, relatively
small ASAS (Pojmañski 1997), MASTER (Lipunov \etal 2010) and new
photopolarimetric system with high temporal resolution 
MiniMegaTORTORA - MMT (Beskin \etal 2013, Biryukov \etal 2015).

The photometric datasets of ground-based surveys are highly affected
by atmospheric extinction. Since the atmosphere is not stationary on
time-scales as short as 0.001--1~s (\eg Dravins \etal 1997), it
influences significantly the observed fast variability of astronomical
sources. The study of some rapid, irregular phenomena like star
flares, transients, occultations etc., becomes difficult, which makes
the accurate atmospheric extinction reduction important.

Many methods of dealing with atmospheric extinction have been suggested 
so far (\eg Straizys 1992), but they typically require subsequent
observations of one or more sources that usually takes a longer time 
than atmospheric non-stationarity scale and as a result an additional 
error is introduced to the de-extincted photometric magnitudes. 
For observations with high temporal resolution one needs a new, improved 
de-extinction procedure based on data obtained in a sufficiently short time. 

There are several catalogs that contain reduced
(``extra-atmospheric'', de-extin\-cted) magnitudes of a large number of
stars. They are based either on measurements obtained during an
orbital mission (like Hipparcos) or on calculations using large number
of observations and monitoring of the local atmosphere (\eg catalog of
bright stars by Kroussanova \etal 2013). Such {\it a priori}
information, as well as a list of photometric standards measured in
various bands, can greatly simplify the calculation of
extra-atmospheric magnitudes and, finally, provide a method for rapid,
but accurate, ground-based photometry.

In this paper we propose a prospective self-consistent method for
atmospheric extinction reduction and a study of local atmosphere
within wide-field multicolor sky observations using extra-atmospheric
magnitudes of photometric standards.

\section{Method Description}
Our method requires that two conditions are met. First, the
observations of the source (or field) in different photometric bands
have to be undertaken (quasi) simultaneously. It is sufficient if the
time interval between exposures taken in different bands is shorter
than the typical atmospheric instability time. It can be achieved
through either simultaneous observations in several bands by
independent telescopes (as implemented in the MMT, see Section~3 for
details) or a series of rather short exposures made by the same
instrument.

The second condition is the availability of a list of photometric
standards with known extra-atmospheric magnitudes in the same bands as
used in the observations. The instrumental photometric system,
however, does not have to coincide with that of the catalog of
standards. (The instruments are always affected by the external
conditions like air temperature, pressure etc., anyway.) If the
spectral transmission curves for all bands of both (``instrumental''
and ``standard'') systems are known, it is possible to convert the
apparent magnitudes of stars from one system to another.

The relationship between the magnitudes in both systems can be
presented in the form of widely-used photometric (color) polynomials.
The calculation of their coefficients is a well-known though quite
complicated procedure. It requires the knowledge of ``typical" spectra
of stars (patterns) for different spectral and luminosity classes (\eg
Pickles 1998) and interstellar medium transmitting spectral curve (\eg
Fitzpatrick 1999). All of these data easily can be found in the
literature. Using such converting polynomials, a catalog of
extra-atmospheric magnitudes of standard stars in the instrumental
photometric system can be built for any telescope and detector.

The atmospheric extinction reduction procedure, used in our method is
as follows. Let $m_0$ be an extra-atmospheric magnitude of an observed
photometric standard in one of instrumental photometric bands, while
$m$ -- is its actually observed (ground-based) magnitude in the same
band. For the star observed at the air mass $M(z)$ (where $z$ is the
zenith distance):
$$m=m_0+a_m\cdot M(z)+C_m\eqno(1)$$
where $a_m$ is an atmospheric extinction coefficient for the
instrumental color band $m$, and $C_m$ is a parameter, which
characterizes the current telescope (and detector) state and does not
depend on the azimuthal coordinates of the star. Furthermore, let
$$a_{m, 0}=(m-m_0)\mid_{M(z)=1}\eqno(2)$$
be the extinction coefficients, calculated for each standard star
assuming some known atmospheric extinction model. Such a model has to
be calculated independently and appears as an initial approximation of
the real state of the atmosphere.

Thus, Eq.(1) can be rewritten as:
$$m=m_0+[a_{m,0}+\Delta a_m]\cdot M(z)+C_m\eqno(3)$$
where $\Delta a_m$ is an unknown correction to the initially assumed
atmospheric extinction model for the current observational set. Note
that coefficients $a_m$ generally depend on the extra-atmospheric
spectrum of the standard, so:
$$\Delta a_m(CI_{i,0})=\sum_i\sum_{k=0}^{k_{\rm max}}c_{ik}CI_{i,0}^k\eqno(4)$$
where $CI_{i,0}$ are extra-atmospheric color indices within the
instrumental photometric system and $c_{ik}$ are coefficients.

The value of $m$ in Eq.(3) is known directly from the observations,
the values of $m_0$ and $a_{m,0}$ are precalculated and air mass
$M(z)\approx\sec z$ depends on the zenith distance in a common
manner. The solution of the system of Eqs.(3) written for all
photometric standards observed simultaneously in the same field and
various color bands, is the set of values of corrections $\Delta
a_m$. More precisely, this solution contains the parameters $c_{ik}$
-- \ie the dependence of $\Delta a_m$ on the colors of standards. The
instrumental parameter $C_m$ is also a part of the full solution of
Eqs.(3). This shows that the multiband simultaneous observations of
standard stars allow us to obtain parameters of a current real state of
the atmospheric extinction and observational equipment.

Finally, since values of $\Delta a_m(CI_{i,0})$ and $C_m$ are now
known for different instrumental color bands, one can calculate the
extra-atmospheric magnitudes $m_0$ of other (non-standard) field
stars. For a star with observed magnitudes $m$ and color indices
$CI_i$:
$$m_0=m-[a_{m,0}(CI_{i,0})+\Delta a_m(CI_{i,0})]\cdot M(z)-C_m\eqno(5)$$
where polynomials $a_{m,0}(CI_{i,0})$ are precalculated with the same
method as was used above for $a_m$ computing.
 
Note that extra-atmospheric color indices $CI_{i,0}$ are initially
unknown for the observed stars (with the exception of photometric
standards), but due to the smallness of the correction $\Delta a_m$
one can use an iterative procedure to estimate $m_0$. Let us assume
that $CI_{i, 0}\equiv CI_i$ at the first step. The value of $m_0$
obtained in this approximation should be used to calculate the next
approximation of extra-atmospheric colors of the star $CI_{i,0}$.
Iterations have to be continued until the set of $m_0$ for different
color bands corresponds to $\Delta a_m$.

\Section{Method Implementation}
The method described above has been developed as a part of data reduction
procedure for MMT photopolarimetric system.

MMT is a complex of 9 robotic telescopes, which is able to provide
simultaneous {\it b$\vv$r}\footnote{Within this paper we will mark
intrinsic color bands of MMT by small characters: {\it b} -- for blue
band, $\vv$ -- for visible band, and {\it r} -- for red band.}
wide-field observations of stars with $\vv$ up to 11~mag for 0.1~s
exposure. The MMT {\it b$\vv$r} bands are very close to those of classical
Johnson-Cousins photometric bands (Johnson and Morgan 1953, Bessel
1990): $b\sim B$, $\vv\sim V$ and $r\sim R$. The response curves of
both systems are shown in Fig.~1. We use solid lines for the standard
{\it BVR} system and dashed lines for the MMT {\it b$\vv$r} filters.
\begin{figure}[p]
  \centerline{\includegraphics[width=5.5cm, angle=270]{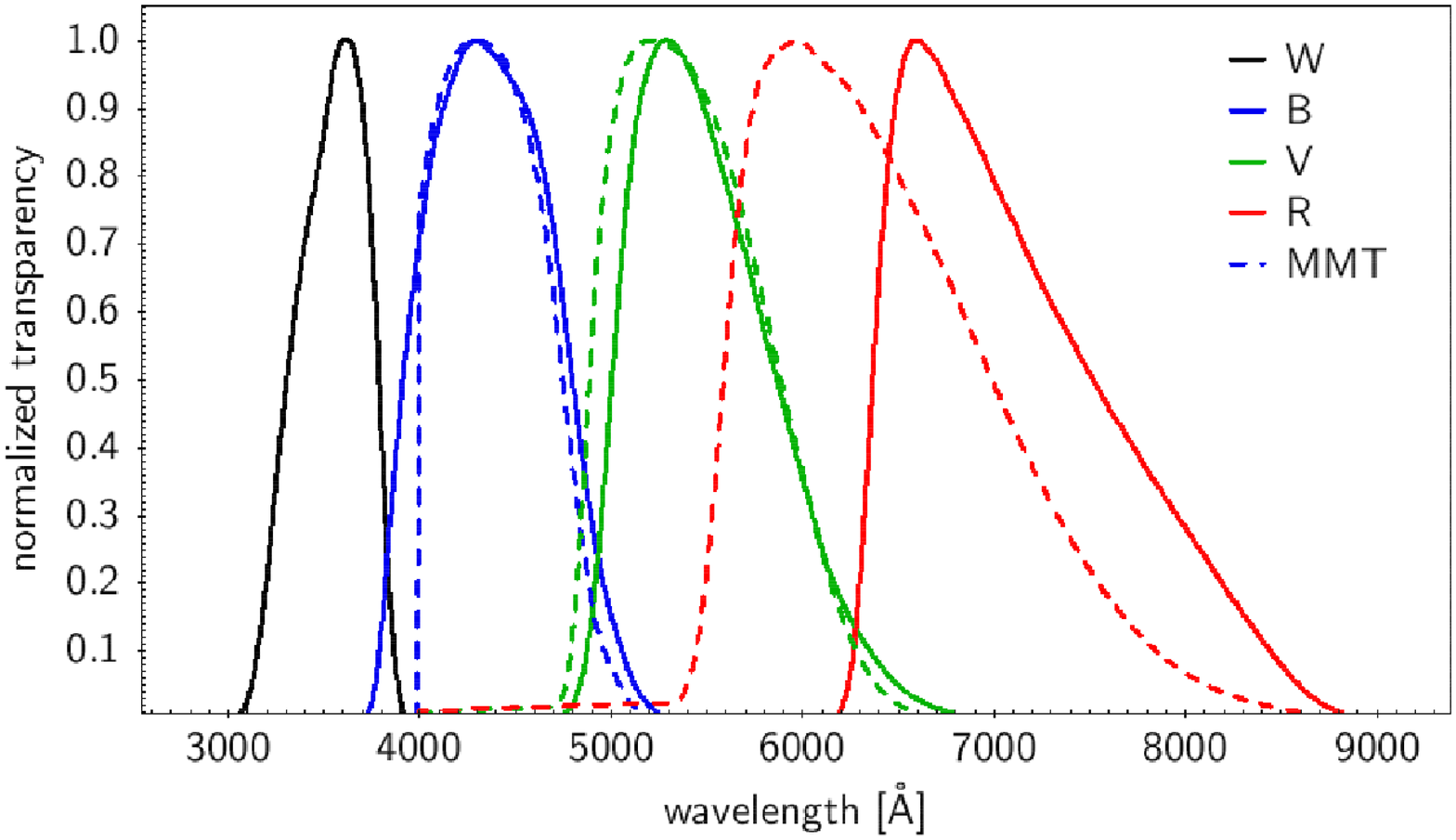}} \vskip3pt
  \FigCap{Spectral response of filters {\it W, B, V} and {\it R} of
    Alma-Ata photometric system (continuous lines) and {\it b$\vv$r}
    filters of MMT complex (dashed lines). The MMT photometric system
    consists of three filters that are close respectively to
    Johnson-Cousins {\it B, V} and {\it R}. The {\it W}-band of Alma-Ata
    system was initially introduced by Straizys (see Straizys 1999 for
    review) as alternative to Johnson's {\it U}. Being revised, this band
    excludes the high influence of the Balmer jump on the measured
    ultra-violet magnitude.}
  \centerline{\includegraphics[width=9.5cm]{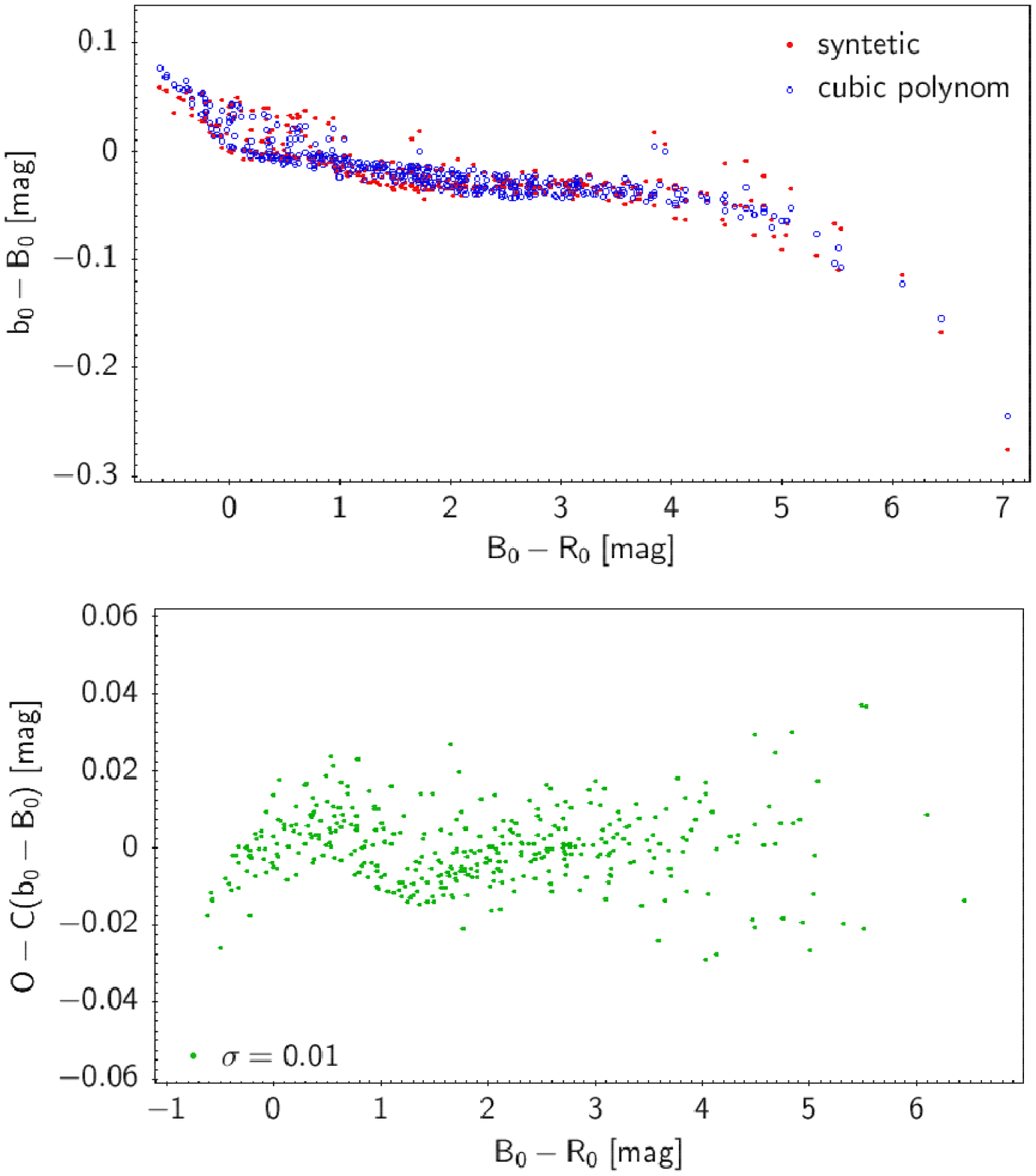}} \vskip3pt
  \FigCap{{\it Upper panel:} the difference between de-extincted
    (extra-atmospheric) magnitude $b_0$ (in MMT photometric system) and
    $B_0$ (in {\it WBVR} system) -- calculated directly from synthetic
    spectra (red points) and using obtained cubic photometric polynomial
    (blue points).  {\it Lower panel:} residuals for $b_0-B_0$ corrections
    obtained with direct calculations and photometric polynomial.}
\end{figure}

Each channel (separate telescope) of MMT is equipped with a standard
Canon 70~mm lens and Neo sCMOS 5Mpix detector from Andor
Technology. An individual channel field of view size is about 100
square degrees, while the whole system's FOV within monitoring regime
is about 900 square degrees. The detection limit (${\rm S/N}=5$) in
the panchromatic band $B=12.0$~mag in 0.1~s (14.5~mag and 17~mag in 10~s
and 1000~s, respectively). In the narrow-field regime of follow-up
observations of individual objects, the size of the field of view
decreases to 100~square degrees, and the detection limit, which depends on
the combination of color and polarization filters, falls within the
range of 10.5--13.5~mag in 0.1~s and reaches 18~mag in 1000~s.

The MMT system collects up to 30Tb of raw data every night. Only an
automatic reduction procedure makes possible the analysis of
accumulated data. Thus, the specific software has been developed for
rapid classification, astrometry and photometry of the observed
phenomena. The photometric module needs a highly efficient method for
atmospheric twinkling reduction -- like that described in Section~2.

Within photometric calibration of MMT data, the catalog of photometric
standards obtained in $WBVR$ Alma-Ata 4-band system (Kroussanova \etal
2013) is used. It contains $\approx 6500$ stars of the northern sky
($\delta>-15\arcd$) with $V\approx6\div7$~mag. The review of the {\it WBVR}
photometric system and the corresponding sky survey can be found in
Kornilov \etal (1991, 1996) and Kornilov (1998). The main advantage of this
catalog is that it contains already de-extincted stellar magnitudes of
stars more or less isotropically distributed over the northern sky.  This
catalog can be used as a basis for automatic reduction of MMT wide-field
observations.

\begin{figure}[htb]
\centerline{\includegraphics[width=7.5cm, angle=270]{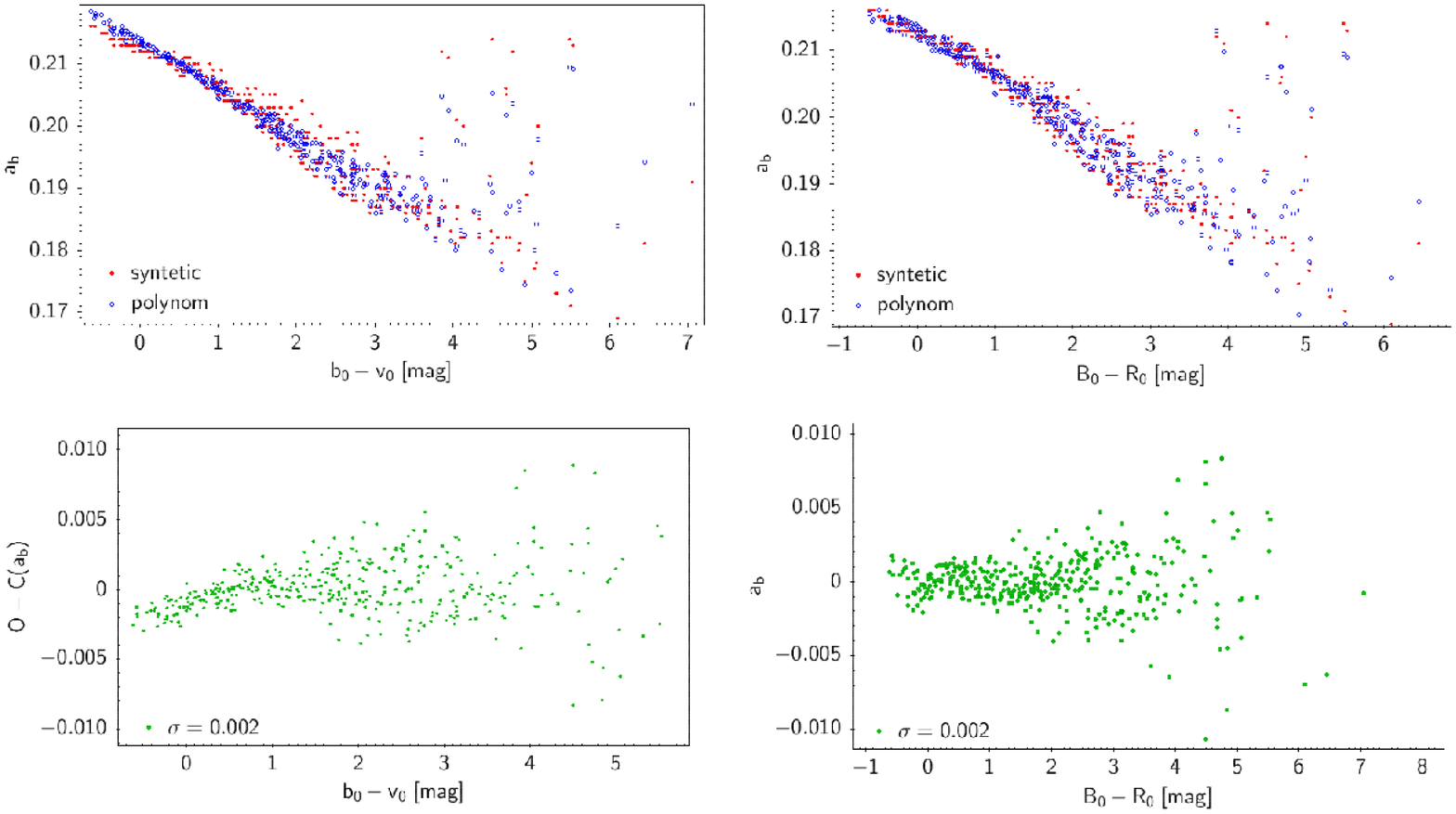}}
\vskip6pt
\FigCap{Atmospheric extinction coefficients $a_b$ for 
MMT band $b$, calculated directly from synthetic spectra
(red points) and using obtained cubic photometrical polynomial (blue
points).}
\end{figure}
Using the atlas of synthetic spectra compiled by Pickles (1998), the
interstellar extinction model provided by Fitzpatrick (1999) and the
model of atmospheric extinction which is relatively close to that of
the MMT site,\footnote{A. Mironov, private
communication} we have calculated the photometric polynomials that
provide the relation between de-extincted $WBVR$ and {\it b$\vv$r}
magnitudes (see Fig.~2), and between atmospheric extinction
coefficients $a_{m,0}$ and color indices $b_0-\vv_0$, $\vv_0-r_0$ (see
Fig.~3). Using these polynomials one can calculate actual de-extincted
magnitudes of the stars observed with MMT and investigate the
atmospheric conditions over the telescope site.

The quite small differences ($\approx0.1-0.2$~mag) between ``real''
(synthetic) and calculated de-extincted magnitudes and colors of the
stars represent the expected accuracy of the described method.

\Section{Discussion and Conclusions}
High precision photometry with accurate atmospheric extinction modeling is
crucial for investigating various astrophysical phenomena accompanied by
fast optical variability. Non-stationary processes in the atmospheres of
Sun-like stars and the main sequence red dwarfs may serve as an
example. These objects are important in the context of the search for
earth-like exoplanets located in habitable zones (Kasting \etal 1993,
Quintana \etal 2014). The observed activity of such stars is partially due
to spots on their surfaces (Gershberg 2005, Maehara \etal 2012) which
affects the locations and widths of their habitable zones. To investigate
this effect, the long-term optical monitoring of a large subset of the main
sequence stars is needed. Such monitoring should be performed with a high
temporal resolution and precise photometry in different color bands. This
is one of the main tasks of the MMT system, and the method described in
this paper is expected to substantially improve the efficiency of such
observations.

We have proposed a new, self-consistent method of atmospheric extinction
reduction and the study of local atmosphere within wide-field multicolor
sky observations using {\it a priori} information about extra-atmospheric
magnitudes of photometric standards. We have also calculated coefficients
of cubic photometric polynomials which are necessary for the implementation
of this method for the new photopolarimetric wide-field telescope system
MMT (Biryukov \etal 2015). We conclude that the calculated synthetic
polynomials are able to provide the atmospheric correction with the
accuracy of the order of 0.1--0.2~mag. In the real observations this
accuracy will slightly decrease and will highly be dependent on the signal-to-noise
ratio for observed stars. For a sufficiently high S/N the final accuracy of
stellar de-extincted photometry will be close to that of the method. In any
case, the 0.1--0.2~mag precision for high temporal resolution data (the
typical MMT exposure is just 0.1~s) can be considered as a major
breakthrough in the stellar photometry.

\Acknow{The work is performed according to the Russian Government 
Program of Competitive Growth of Kazan Federal University.}

\end{document}